\documentclass[11pt]{article}

\usepackage{fullpage}
\usepackage{cite}
\usepackage{graphicx}
\usepackage{hyperref}
\usepackage{amssymb,comment,algorithm,algorithmic,bbm,stmaryrd}
\usepackage{amsmath,amsfonts,amssymb,amsthm,ifsym}
\usepackage{multirow,rotating}
\usepackage{multicol}
\usepackage{pifont}
\usepackage[usenames]{color}

\newcommand{\HH}{\mathcal{H}}
\newcommand{\DD}{\mathcal{D}}
\newcommand{\paren}[1]{\left( #1 \right)}
\newcommand{\abs}[1]{\left| #1 \right|}
\newcommand{\cut}[1]{}
\newcounter{rmk}

\newenvironment{reminderlemma}[1]{\medskip

\noindent {\bf Reminder of Lemma #1.  }\em}{}
\newtheorem{remark}[rmk]{Remark}
\def \QED {\hfill{$\Box$}}
\DeclareMathAlphabet {\mathpzc}{OT1}{pzc}{m}{it}

\newenvironment{proofof}[1]{\noindent {\em Proof of #1.  }}{\QED}
\newtheorem{theorem}{Theorem}

\newtheorem{claim}[theorem]{Claim}
\newtheorem{lemma}[theorem]{Lemma}

\newtheorem{definition}[theorem]{Definition}
\newenvironment{reminderclaim}[1]{\medskip
\noindent {\bf Reminder of Claim #1.  }\em}{}

\title{Differentially Private Data Analysis of Social Networks via Restricted Sensitivity}

\author{Jeremiah Blocki \and Avrim Blum \and Anupam Datta \and Or Sheffet\\
{\small Carnegie Mellon University}\\
\texttt{ \{jblocki@cs, avrim@cs, danupam@andrew, osheffet@cs\}.cmu.edu }\thanks{This research was supported in part by CyLab at Carnegie Mellon University under grants DAAD19-02-1-0389 and W911NF-09-1-0273 from the Army Research Office, the National Science Foundation under grants CCF-1101215 and CCF-1116892, the NSF Science and Technology Center TRUST,
and an NSF Graduate Fellowship, as well as by the MSR-CMU Center for
Computational Thinking.
}
}
\date{\today \\}

\begin{document}
\begin{titlepage}

\maketitle

\begin{abstract}
We introduce the notion of {\em restricted sensitivity} as an
alternative to global and smooth sensitivity to improve accuracy in
differentially private data analysis.  The definition of restricted
sensitivity is similar to that of global sensitivity except that
instead of quantifying over all possible datasets, we take advantage
of any beliefs about the dataset that a querier may have, to
quantify over a restricted class of datasets.  Specifically, given a
query $f$ and a hypothesis $\HH$ about the structure of a dataset
$D$, we show generically how to transform $f$ into a new
query $f_\HH$ whose global sensitivity (over all datasets including
those that do not satisfy $\HH$) matches the restricted sensitivity of
the query $f$.  Moreover, if the belief of the querier is correct
(i.e., $D \in \HH$) then $f_\HH(D) = f(D)$.  If the belief is incorrect, then
$f_\HH(D)$ may be inaccurate.

We demonstrate the usefulness of this notion by considering the task
of answering queries regarding social-networks, which we model as a
combination of a graph and a labeling of its vertices.  In particular,
while our generic procedure is computationally inefficient, for the
specific definition of $\mathcal{H}$ as graphs of bounded degree, we
exhibit efficient ways of constructing $f_{\cal H}$ using different
projection-based techniques. We then analyze two important query
classes: \emph{subgraph counting queries} (e.g., number of triangles)
and \emph{local profile queries} (e.g., number of people who know a
spy and a computer-scientist who know each other).  We demonstrate that the restricted
sensitivity of such queries can be significantly lower than their
smooth sensitivity.  Thus, using restricted sensitivity we can
maintain privacy whether or not $D \in \HH$, while providing more
accurate results in the event that $\HH$ holds true.
\end{abstract}

\thispagestyle{empty}
\end{titlepage}

\section{Introduction}

The social networks we inhabit have grown significantly in recent
decades with digital technology enabling the rise of networks like
Facebook that now connect over $900$ million people and house vast
repositories of personal information. At the same time, the study of
various characteristics of social networks has emerged as an active
research area~\cite{easley2010networks}. Yet the fact that the data in
a social network might be used to infer sensitive details about an
individual, like sexual orientation \cite{jernigan2009gaydar}, is a
growing concern among social networks' participants. Even in an
`anonymized' unlabeled graph it is possible to identify people based
on graph structures \cite{backstrom2007wherefore}.  In this paper, we
study the feasibility of and design efficient algorithms to release
statistics about social networks (modeled as graphs with vertices
labeled with attributes) while satisfying the semantic definition of
differential privacy~\cite{dwork2006differential,Dwork06calibratingnoise}.

A differentially private mechanism guarantees that any two neighboring
data sets (i.e., data sets that differ only on the information about a
single individual) induce similar distributions over the statistics
released.  For social networks, we consider two notions of neighboring
or adjacent networks: (1) \emph{edge adjacency} stipulating that
adjacent graphs differ in just one edge or in the attributes of just one
vertex; and (2) \emph{vertex adjacency} stipulating that adjacent
networks differ on just one vertex---its attributes or {\em any number} of
edges incident to it.

For any given statistic or query, its global sensitivity
measures the maximum difference in the answer to that query over all pairs of 
neighboring data sets~\cite{Dwork06calibratingnoise}; global 
sensitivity provides an upper bound on the amount of noise that has to be added to the actual 
statistic in order to preserve differential privacy. Since the global sensitivity 
of certain types of queries can be quite high, the notion of smooth sensitivity was 
introduced to reduce the amount of noise that needs to be added while still 
preserving differential privacy~\cite{nissim2007smooth}.
 
However, a key challenge in the differentially private analysis of
social networks is that for many natural queries, both global and
smooth sensitivity can be very large.  In the vertex adjacency model,
consider the query ``How many people in $G_1$ are a doctor or are
friends with a doctor?"  Even if the answer is $0$ (e.g., there are no
doctors in the social network) there is a neighboring social network
$G_2$ in which the answer is $n$ (e.g., pick an arbitrary person from
$G_1$, relabel him as a doctor, and connect him to everyone). Even in
the edge adjacency model, the sensitivity of queries may be
high. Consider the query ``How many people in $G_1$ are friends with
two doctors who are also friends with each other?" In $G_1$ the answer
may be $0$ even if there are two doctors that everyone else is friends
with (e.g, the doctors are not friends with each other), but the
answer jumps to $n-2$ in a neighboring graph $G_2$ (e.g, if we simply
connect the doctors to each other).  In fact, even the first query can
have high sensitivity in the edge-adjacency model if we just relabel a
high-degree vertex as a doctor.

Yet, while these examples respect the mathematical definitions of
neighboring graphs and networks, we note that in a real social network
no single individual is likely to be directly connected with everyone
else.  Suppose that in fact a querier has some such belief $\HH$ about
the given network ($\HH$ is a subset of all possible networks) such
that its query $f$ has low sensitivity restricted only to inputs and
deviations within $\HH$.  For example, the querier may believe the
following hypothesis $(\HH_k)$: the maximum degree of any node in the
network is at most $k=5000 \ll n \approx 9\times 10^8$ (e.g, after
reading a study on the anatomy of Facebook \cite{ugander2011anatomy}).
Can one in that case provide accurate answers in the event that indeed
$G \in \HH$ and yet preserve privacy no matter what (even if $\HH$ is
not satisfied)?   

In this work, we provide a positive answer to this question.  We do so
by introducing the notion of \emph{restricted sensitivity}, which
represents the sensitivity of the query $f$ over only the given subset
$\HH$, and providing procedures that map a query $f$ to an alternative
query $f_{\cal H}$ s.t. $f$ and $f_{\cal H}$ identify over the inputs
in ${\cal H}$, yet the global sensitivity of $f_{\cal H}$ is
comparable to just the restricted sensitivity of $f$.  Therefore, the
mechanism that answers according to $f_{\cal H}$ and adds Laplace
random noise preserves privacy for \emph{all} inputs, while giving
good estimations of $f$ for inputs in $\HH$.

While our general scheme for devising such $f_{\cal H}$ is
inefficient and requires that we construct a separate $f_{\cal H}$
for each query $f$, we also design a complementary
\emph{projection-based} approach. A projection of $\mathcal{H}$ is a
function mapping all possible inputs (e.g., all possible $n$-node
social networks) to inputs in $\mathcal{H}$ with the property that any
input in $\mathcal{H}$ is mapped to itself. Therefore, a projection
$\mu$ allows us to define $f_{\cal H}$ for any $f$, simply by
composing $f_{\cal H} = f\circ \mu$.  Moreover, if this projection
$\mu$ satisfies certain smoothness properties, which we define in
Section~\ref{sec:Results}, then this function $f_\HH$ will have its
global sensitivity---or at least its smooth sensitivity over inputs
in $\HH$---comparable to only the restricted sensitivity of $f$.  In
particular, for the case $\HH = \HH_k$ (the assumption that the
network has degree at most $k \ll n$), we show we can {\em
efficiently} construct projections $\mu$ satisfying these conditions,
therefore allowing us to efficiently take advantage of low restricted
sensitivity.  These results are given in Section \ref{sec:Results} and
summarized in Table \ref{table:Results}.

The next natural question is: how much advantage does restricted
sensitivity provide, compared to global or smooth sensitivity, for
natural query classes and natural sets $\HH$?  In Section
\ref{sec:RestrictedSensitivity} we consider two natural classes of
queries: local profile queries and subgraph counting queries. A local
profile query asks how many nodes $v$ in a graph satisfy a property
which depends only on the immediate neighborhood of $v$ (e.g, queries
relating to clustering coefficients and
bridges~\cite{easley2010networks}, or queries such as ``how many
people know two spies who don't know each other?'').  A subgraph
counting query asks how many copies of a particular subgraph $P$ are
contained in the network (e.g., number of triangles involving at least
one spy).  For the case $\HH = \HH_k$ for $k \ll n$ we show that the
restricted sensitivity of these classes of queries can indeed be much
lower than the smooth sensitivity.  These results, presented in
Section \ref{sec:RestrictedSensitivity}, are summarized in Table
\ref{table:NoiseLevel}.

\begin{table*}
\centering
  \begin{tabular}{ | l || c | c | c | c | c | }
\hline  & Adjacency & Hypothesis & Query  & Sensitivity   & Efficient   \\ \hline Theorem
\ref{thm:ConstructFprime} & Any & Any & Any & $GS_{f_{\cal H}} = RS_f\paren{\cal H}$ & No  \\ \hline Theorem
\ref{thm:ConstructFprimeEfficient} & Edge  & ${\cal H}_k$ & Any & $GS_{f_{\cal H}} = 3 RS_f\paren{\cal H}$ & Yes  \\ \hline
Theorem \ref{thm:main} & Vertex  & ${\cal H}_k$ & Any & $S_{f_{\cal H}} = O\paren{1}\times RS_f\paren{{\cal H}_{2k}}$ & Yes  \\ \hline
\end{tabular} \caption{Summary of Results.  $GS = $ global sensitivity, 
$RS = $ restricted sensitivity, and $S = $ smooth bound of local sensitivity. 
\label{table:Results}} 
\end{table*}

\begin{table*}
\centering
\begin{tabular}{ | l || p{.7in} |
p{.7in} | p{.7in} | p{.7in} | } \hline  & \multicolumn{2}{c|}{Subgraph Counting
Query $P$} &       \multicolumn{2}{c|}{Local Profile Query}  \\ \hline
Adjacency  & Smooth  & Restricted    & Smooth  & Restricted  \\ \hline 
Edge  & ${|P|~k^{|P|-1} }$ & ${|P|~k^{|P|-1}}$ & $k+1$ & $k+1$ \\ \hline Vertex & $O\paren{n^{|P|-1}}$ & ${|P|~k^{|P|-1}}$ & $n-1$ & $2k+1$ \\ \hline
\end{tabular} \caption{ Worst Case Smooth Sensitivity over ${\cal H}_k$ vs. Restricted Sensitivity $RS_f\paren{{\cal H}_k}$.
\label{table:NoiseLevel}}
\vspace{-.2in}
\end{table*}

\subsection{Related Work} \label{subsec:related}
Easley and Kleinberg provide an excellent summary of the rich literature on social networks \cite{easley2010networks}. Previous literature on differentially-private analysis of social networks has primarily focused on the edge adjacency model in unlabeled graphs where sensitivity is manageable \footnote{Kasiviswanathan, Nissim, Raskhodnikova and Smith have independently been exploring and developed an analysis for node level privacy using an  approach similar to ours (personal communication, 2012).}.  Triangle counting queries can be answered in the edge adjacency model by efficiently computing the smooth sensitivity~\cite{nissim2007smooth}, and this result can be extended to answer other counting queries \cite{KRSY11}. \cite{HayLMJ09} shows how to privately approximate the degree distribution in the edge adjacency model. The Johnson-Lindenstrauss transform can be used to answer all cut queries in the edge adjacency model \cite{JLprivacy}. 

The approach taken in the work of Rastogi et al. \cite{rastogi2009relationship} on answering subgraph counting queries is the most similar to ours. They consider a bayesian adversary whose prior (background knowledge) is drawn from a distribution. Leveraging an assumption about the adversary's prior they compute a high probability upper bound on the local sensitivity of the data and then answer by adding noise proportional to that bound. Loosely, they assume that the presence of an edge does not presence of other edges more likely. In the specific context of a social network this assumption is widely believed to be false (e.g., two people are more likely to become friends if they already have common friends \cite{easley2010networks}). The privacy guarantees of \cite{rastogi2009relationship} only hold if these assumptions about the adversaries prior are true. By contrast, we always guarantee privacy even if the assumptions are incorrect.

A relevant approach that deals with preserving differential privacy while providing better utility guarantees for nice instances is detailed in the work of Nissim et al~\cite{nissim2007smooth} who define the notion of smooth sensitivity. In their framework, the amount of random noise that the mechanism adds to a query's true ansewr is dependent on the extent for which the input database is ``nice'' -- having small \emph{local sensitivity}. As we discuss later, in social networks many natural queries (e.g., local profile queries) even have high local and smooth sensitivity.

\section{Preliminaries}
\label{sec:preliminaries}

\subsection{Differential Privacy}
\label{subsec:diff_privacy}

We adopt the framework of differential privacy. We use $\mathcal{D}$ to denote the set of all possible datasets. Intuitively, we say two datasets $D,D' \in \mathcal{D}$ are \emph{neighbors} if they differ on the details of a single individual. (See further discussion in Definitions~\ref{def_Privacy:EdgeChanges} and~\ref{def_Privacy:VertexChanges}.) We denote the fact that $D'$ is a neighbor of $D$ using $D'\sim D$. We define the \emph{distance} $d(D,D')$ between two databases $D,D'\in\mathcal{D}$ as the minimal non-negative integer $k$ s.t. there exists a path $D_0, D_1, \ldots, D_k$ where $D_0 = D$, $D_k = D'$ and for every $1\leq i\leq k$ we have that $D_{i-1} \sim D_{i}$. Given a subset $\mathcal{D}' \subset \mathcal{D}$ we denote the distance of a database $D$ to $\mathcal{D}'$ as $d(D,\mathcal{D}') = \min\limits_{D' \in \mathcal{D'}} d(D, D')$. 

\begin{definition}\cite{dwork2006differential}
A mechanism $A$ is $(\epsilon, \delta)$-differentially private if
for every pair of neighboring datasets $D,D' \in \mathcal{D}$ and every subset $S\subseteq Range(A)$ we have that  $\Pr[A(D) \in S] \leq e^\epsilon \Pr[A(D') \in S] + \delta \ . $ 
\end{definition}

Intuitively differential privacy guarantees that an adversary has a very limited ability to distinguish between the output of $A\paren{D}$ and the output of $A\paren{D'}$. A query is a function $f:\mathcal{D} \rightarrow \mathbb{R}$ mapping the dataset to a real number.

\begin{definition} \label{def:LocalSensitivity}
The {\em local sensitivity} of a query $f$ at a dataset $D$ is $LS_f(D) = \max\limits_{D'\sim D} \abs{f(D)-f(D')}$.
\end{definition}

\begin{definition} \label{def:GlobalSensitivity}
The {\em global sensitivity} of a query $f$ is $GS_f = \max_{D\in \mathcal{D}} LS_f(D)$. 
\end{definition}

The Laplace mechanism $A\paren{D} = f\paren{D} + {Lap}\paren{ GS_f/\epsilon}$ preseves $(\epsilon,0)$-differential privacy \cite{Dwork06calibratingnoise}. This mechanism provides useful answers to queries with low global sensitivity. The primary challenge in the differentially private analysis of social networks is the high global sensitivity of many queries. The local sensitivity $LS_f(D)$ may be significantly lower than the global sensitivity $GS_f$.  However, adding noise proportional to $LS_f(D)$ does not preserve differential privacy because the noise level itself may leak information.  A clever way to circumvent this problem is to smooth out the noise level \cite{nissim2007smooth}.

\begin{definition} \cite{nissim2007smooth} \label{def:BetaSmoothUpperBound}
A $\beta$-smooth upper bound  on the local sensitivity of a query $f$ is a function $S_{f,\beta}$ which satisfies (i) $\forall D \in \mathcal{D}, S_{f,\beta}\paren{D} \geq LS_f\paren{D}$, and (ii) $\forall D,D' \in \mathcal{D}$ it holds that $S_{f,\beta}\paren{D} \leq \exp\paren{-\beta d\paren{D,D'}}S_{f,\beta}\paren{D'}$.
\end{definition}

It is possible to preserve privacy while adding noise proportional to a $\beta$-smooth upper bound on the sensitivity of a query. For example, the mechanism $A\paren{D} = f\paren{D} + Lap\paren{\frac{2S_{f,\beta}\paren{D}}{\epsilon}}  \ ,$
with $\beta = - \epsilon/2 \ln \delta$ preserves $(\epsilon,\delta)$-differential privacy \cite{nissim2007smooth}. To evaluate $A$ efficiently one must present an algorithm to efficiently compute the $\beta$-smooth upperbound $S_{f,\beta}\paren{G}$, a task which is by itself often non-trivial.
%This is not always an easy task! Unfortuanately, even if we find an efficiently computable $\beta$-smooth upperbound $S_{f,\beta}\paren{G}$ for the query $f$ the value of $S_{f,\beta}\paren{G}$ may be prohibatively high in our labeled graph setting. 

\subsection{Graphs and Social Networks}
\label{subsec:graphs}

Our work is motivated by the challenges posed by differentially private analysis of social networks. As always, a graph is a pair of a set of vertices and a set of edges $G = \langle V, E\rangle$. We often just denote a graph as $G$, referring to its vertex-set or edge-set as $V(G)$ or $E(G)$ resp. A key aspect of our work is modeling a social network as a \emph{labeled} graph.
\begin{definition} \label{def:SocialNetwork}
A {\em social network}  $(G,\ell)$ is a graph with labeling function $\ell:V(G) \to \mathbb{R}^m$. The set of all social networks is denoted $\mathcal{G}$.
\end{definition}
The labeling function $\ell$ allows us to encode information about the nodes (e.g., age, gender, occupation). For convenience, we assume all social networks are over the same set of vertices, which is denotes as $V$ and has size $n$, and so we assume $|V|=n$ is public knowledge.\footnote{Adding or removing vertices could be done by adding one more dimension to the labeling, indicating whether a node is active or inactive.} Therefore, the graph structures of two social networks are equal if their edge-sets are identical. 
%\Look{The last two sentences above are confusing - makes the reader think that the labeling function is fixed and public, and that the only thing private is the edges.  Is the point here just that for convenience we assume $n=|V|$ is public?  If so, let's just say that.} 
Similarly, we also fix the dimension $m$ of our labeling.

%A labeled graph could be used to represent a social network like Facebook, a telephone call network, a collaboration network, etc.

Defining differential privacy over the labeled graphs $\mathcal{G}$ requires care. What does it mean for two labeled graphs $G_1, G_2 \in \mathcal{G}$ to be neighbors? There are two natural notions: edge-adjacency and vertex adjacency.

\begin{definition} [Edge-adjacency]
\label{def_Privacy:EdgeChanges} We say that two social networks $(G_1,\ell_1)$ and $(G_2,\ell_2)$ are neighbors if either (i) $E(G_1) = E(G_2)$ and there exists a vertex $u$ such that $\ell_1(u)\neq \ell_2(u)$ whereas for every other $v\neq u$ we have $\ell_1\paren{v} = \ell_2\paren{v}$ or (ii) $\forall v, \ell_1(v) = \ell_2\paren{v}$ and the symmetric difference $E(G_1) \vartriangle  E(G_2)$ contains a single edge.
\end{definition}

In the context of a social network, differential-privacy w.r.t edge-adjacency can, for instance, guarantee that an adversary will not be able to distinguish whether a particular individual has friended some specific pop-singer on Facebook. However, such guarantees do not allow a person to pretend to listen only to high-end indie rock bands, should that person have friended numerous pop-singers on Facebook. This motivates the stronger vertex-adjacency neighborhood model.
%\Look{Can the def below be simplified to just say that $(G_1,\ell_1)$ and $(G_2,\ell_2)$ are neighbors if there exists $u$ such that $G_1 - u = G_2 - u$ and $\ell_1(v) = \ell_2(v)$ for all $v \neq u$?  Or if a vertex wants to change both its attributes and its incident edges, do we count that as distance 2?}

\begin{definition} [Vertex-adjacency] 
\label{def_Privacy:VertexChanges} 
We say that two social networks $(G_1,\ell_1)$ and $(G_2,\ell_1)$ are neighbors if there exists a vertex $v_i$ such that $G_1-v_i = G_2-v_i$ and $\ell_1(v_j)= \ell_2(v_j)$ for every $v_j\neq v_i$.
\end{definition}
\noindent where for a graph $G$ and a vertex $v$ we denote $G-v$ as the result of  removing every edge in $E(G)$ that touches $v$.

It is evident that any two social networks that are edge-adjacent are also vertex-adjacent. Preserving differential privacy while guaranteeing good utility bounds w.r.t vertex-adjacency is a much harder task than w.r.t edge-adjacency.

\paragraph{Distance} Given two social networks $(G_1, \ell_1)$ and $(G_2, \ell_2)$, recall that their distance is the minimal $k$ s.t. one can form a path of length $k$, starting with $(G_1,\ell_1)$ and ending at $(G_2,\ell_2)$, with the property that every two consecutive social-networks on this path are adjacent. Given the above two definitions of adjacency, we would like to give an alternative characterization of this distance. 

First of all, the set $U = \{v:~ \ell_1(v)\neq \ell_2(v)\}$ dictates $|U|$ steps that we must take in order to transition from $(G_1,\ell_1)$ to $(G_2, \ell_2)$. It is left to determine how many adjacent social-networks we need to transition through until we have $E(G_1) = E(G_2)$. To that end, we construct the difference-graph whose edges are the symmetric difference of $E(G_1)$ and $E(G_2$). Clearly, to transition from $(G_1,\ell_1)$ to $(G_2, \ell_2)$, we need to alter every edge in the difference graph. In the edge-adjacency model, a pair of adjacent social networks covers precisely a single edge, and so it is clear that the distance $d\big((G_1,\ell_1), (G_2,\ell_2)\big) = |U| + |E(G_1) \vartriangle  E(G_2)|$. In the vertex-adjacency model, a single vertex can cover all the edges that touch it, and so the distance between the graphs $G_1-U$ and $G_2-U$ is precisely the \emph{vertex cover} of the difference graph. Denoting this vertex cover as $VC(G_1-U\vartriangle  G_2-U)$ we have that $d\big((G_1,\ell_1), (G_2,\ell_2)\big) = |U| + |VC(G_1 - U \vartriangle  G_2-U)|$. It is evident that computing the distance of between any two social-networks in the vertex-adjacency model is a NP-hard problem.

%\paragraph{Subgraphs.} Given $k$ disjoint vertices in $V$, we denote by $G[v_1, v_2,\ldots, v_k]$ the subgraph of $G$ induced by these $k$ vertices and all edges between them, and by $\ell[v_1, v_2, \ldots, v_k]$ the induced labeling of these $k$ vertices. Given a triplet $\langle u, H, \ell'\rangle$ where $(H,\ell')$ is some social-network over $n' \leq n$ vertices and $u \in V(H)$, we say $(G,\ell)$ \emph{induces the rooted social-network $\langle u, H, \ell'\rangle$ at $v$} if there exists a mapping $\phi:V(G)\to V(H)\cup \{\bot\}$ s.t. (i) $\phi(v) = u$, (ii) for every $x'\in V(H)$ there exists a single $x\in V(G)$ s.t. $\phi(x)=x'$, (iii) for every $v$ with $\phi(v) \neq \bot$ we have $\ell'(\phi(v)) = \ell(v)$, and (iv) for every $x,y$ s.t. $\phi(x),\phi(y)\neq \bot$ we have $(x,y)\in E(G)$ iff $(\phi(x),\phi(y))\in E(H)$.

To avoid cumbersome notation, from this point on we omit the differentiation between graphs and social networks, and denote networks as graphs $G \in \mathcal{G}$. 

%\Look{Move to Section \ref{sec:RestrictedSensitivity}, or remove if we don't use it.}
%We say that the $2$-betweeness of a vertex $v$ is the probability that the a randomly chosen shortest path between two randomly chosen neighbors $x,y \in N_v$ goes through $v$. Betweeness \cite{freeman1979centrality} can also be used to measure the influence of a node $x$.

%\input{itcs_definitions}
\section{Restricted Sensitivity}
 \label{sec:restriced_sensitivity} 

We now introduce the notion of restricted sensitivity, using a hypothesis about the dataset $D$ to restrict the sensitivity of a query. A hypothesis $\mathcal{ {\cal H}}$ is a subset of the set $\DD$ of all possible datasets (so in the context of social networks, $\HH$ is a set of labeled graphs).  We say that ${\cal {\cal H}}$ is true if the true dataset $D \in {\cal {\cal H}}$. Because the hypothesis ${\cal {\cal H}}$ may not be a convex set we must consider all pairs of datasets in ${\cal {\cal H}}$ instead of all pairs of adjacent datasets as in the definition of global sensitivity.

\begin{definition} \label{def:RestrictedSensitivity}
For a given notion of adjacency among datasets, 
the {\em restricted sensitivity} of $f$ over a hypothesis ${\cal {\cal H}} \subset \mathcal{D}$ is 
\[ RS_f\paren{{\cal {\cal H}}} = \max_{D_1,D_2 \in {\cal {\cal H}}} \big(\frac{\abs{f\paren{D_1}-f\paren{D_2}}}{d\paren{D_1,D_2}} \big) \ .\] 
\end{definition}
To be clear, $d(D_1,D_2)$ denotes the length of the shortest-path in $\DD$ between $D_1$ and $D_2$  (not restricting the path to only use $D \in \HH$) using the given notion of adjacency (e.g., edge-adjacency or vertex-adjacency).
%%We clarify: the distance in the denominator in Definition~\ref{def:RestrictedSensitivity} is precisely the same distance defined in the Preliminaries. 
That is, we restrict the set of databases for which we compute the sensitivity, but we do not re-define the distances. 
%%For any two $D_1, D_2\in\mathcal{H}$ the distance is still the shortest path of adjacent databases (in $\mathcal{D}$, and not necessarily in $\mathcal{H}$) connecting $D_1$ with $D_2$.

Observe that $RS_f\paren{{\cal {\cal H}}}$ may be smaller than $LS_f\paren{D}$ for some $D \in {\cal {\cal H}}$ if $D$ has a neighbor $D' \notin {\cal {\cal H}}$. In fact we often have $LS_f\paren{D} \geq \left| f\paren{D} -f\paren{D'} \right| \gg RS_f\paren{{\cal {\cal H}}}$.
As an immediate corollary, in such cases $RS_f\paren{{\cal {\cal H}}}$ will be significantly lower than $S_{f,\beta}\paren{D}$, a $\beta$-smooth upper bound on $LS_f\paren{D}$.

%In general the mechanism $A\paren{D} = f\paren{D} + Lap\paren{RS_f\paren{{\cal {\cal H}}}/\epsilon}$ does not preserve differential privacy. 

%%To achieve differential privacy while adding noise proportional to $RS_f\paren{{\cal {\cal H}}}$ we must be willing to sacrifice accuracy guarantees for datasets $D \notin {\cal {\cal H}}$. Our goal is to create a new query $f_{\cal {\cal H}}$ such that  $f_{\cal {\cal H}}(D) = f(D)$ for every $D \in \mathcal{D}$ ($f_{\cal {\cal H}}$ is accurate when the hypothesis is correct) and $f_{\cal {\cal H}}$ either has low global sensitivity or low $\beta$-smooth sensitivity over datasets $D \in \mathcal{D}$. 
%If we want an efficient mechanism then we must also be able to compute the smoothed sensitivity of $f_{\cal {\cal H}}$ efficiently. 
\cut{
Given a query $f$ and a hypothesis $\HH$, our goal is now to produce an answer to $f$ such that privacy is maintained whether or not $D \in \HH$, but if indeed $D \in \HH$ then noise is proportional only to the restricted sensitivity of $f$ over $\HH$.   We give procedures for this below in Section \ref{sec:Results}, including a general but computationally inefficient procedure that applies to any $\HH$ and any adjacency notion, as well as efficient procedures for important special cases in the context of social networks.  We then in Section \ref{sec:RestrictedSensitivity} compare restricted and smoothed sensitivity for natural classes of queries over social networks when $\HH$ is the low-degree hypothesis.}

\section{Using Restricted Sensitivity to Reduce Noise}
\label{sec:Results}
To achieve differential privacy while adding noise proportional to $RS_f\paren{{\cal {\cal H}}}$ we must be willing to sacrifice accuracy guarantees for datasets $D \notin {\cal {\cal H}}$. Our goal is to create a new query $f_{\cal {\cal H}}$ such that  $f_{\cal {\cal H}}(D) = f(D)$ for every $D \in \mathcal{H}$ ($f_{{\cal H}}$ is accurate when the hypothesis is correct) and $f_{\cal {\cal H}}$ either has low global sensitivity or low $\beta$-smooth sensitivity over datasets $D \in \mathcal{H}$. 
In this section,
we first give a non-efficient generic construction of such $f_{\cal {\cal H}}$, showing that it is always possible to devise $f_{\cal H}$ whose global sensitivity equals \emph{exactly} the restricted sensitivity of $f$ over $\mathcal{H}$.
We then show how for the case of social networks and for the hypothesis $\HH_k$ that the network has bounded degree, we can construct functions $f_{\HH_k}$ having approximately this property, efficiently.

\subsection{A General Construction}
\label{sec:general}

We now show how given $\HH$ to generically (but not efficiently)  construct $f_{\cal H}$ whose global sensitivity \emph{exactly} equals the restricted sensitivity of $f$ over $\mathcal{H}$.

\begin{theorem} \label{thm:ConstructFprime}
Given any query $f$ and any hypothesis ${\cal H} \subset \mathcal{D}$ we can construct a query $f_{\cal H}$ such that 
%\begin{multicols}{2}
\begin{enumerate}
\item $\forall D \in {\cal H}$ it holds that $f_{\cal H}\paren{D} = f\paren{D}$, and
\item $GS_{f_{\cal H}} = RS_f\paren{{\cal H}}$
\end{enumerate}
%\end{multicols}
\end{theorem}
\begin{proof}
For each $D \in {\cal H}$ set $f_{\cal H}\paren{D} = f\paren{D}$. Now fix an arbitrary ordering of the set $\{D :\ D \notin {\cal H}\}$, and denote its elements as $D_1, D_2, \ldots, D_m$, where $m$ is the size of the set. For every $D\notin \mathcal{H}$ we define the value of $f_{\cal H}(D)$ inductively. Denote $\mathcal{T}_i = {\cal H} \bigcup \{D_1,...,D_i\}$. Initially, we are given the values of every $D \in \mathcal{T}_0$. Given $i > 0$, we denote $\Delta_i = RS_{f_{\cal H}}\paren{{\cal T}_i}$.
We now prove one can pick the value $f_{\cal H}(D_{i})$ in a way that preserves the invariant that $\Delta_{i+1} = \Delta_i$. By applying the induction $m$ times we conclude that \[RS_f\paren{{\cal H}} = \Delta_0 = \Delta_m =  RS_{f_{\cal H}}\paren{{\cal D}} = GS_{f_{\cal H}}\ .\]

Fix $i>0$. Observe that \[\Delta_{i+1} = \max \paren{\Delta_i, \paren{ \max_{D \in {\cal T}_i } \frac{\abs{f_{\cal H}\paren{D} -f_{\cal H}\paren{D_{i+1}} }}{d\paren{D,D_{i+1}}  }}}\] 
so to preserve the invariant it suffices to find any value of $f_{\cal H}\paren{D_{i+1}}$ that satisfies that for every $D\in {\cal T}_i$ we have
$\abs{f_{\cal H}\paren{D} - f_{\cal H}\paren{D_{i+1}}} \leq \Delta_i \cdot {d\paren{D,D_{i+1}}}$.
Suppose for contradiction that no value exists. Then there must be two intervals
\begin{align*}
&\left[f_{\cal H}\paren{D_1^*}- \Delta_i\cdot d\paren{D_1^*,D_{i+1}} , \ f_{\cal H}\paren{D_1^*}+\Delta_i\cdot d\paren{D_1^*,D_{i+1}}  \right] \cr 
& \left[f_{\cal H}\paren{D_2^*}-\Delta_i\cdot d\paren{D_2^*,D_{i+1}}, \ f_{\cal H}\paren{D_2^*}+\Delta_i\cdot d\paren{D_2^*,D_{i+1}} \right]\end{align*}  
which don't intersect. This would imply that 
\[\frac{\abs{f_{\cal H}\paren{D_1^*} - f_{\cal H}\paren{D_2^*}}}{d\paren{D_1^*,D_2^*} } \geq \frac{\abs{f_{\cal H}\paren{D_1^*} - f_{\cal H}\paren{D_2^*}}}{d\paren{D_{i+1},D_1^*} + d\paren{D_{i+1},D_2^*}} > \Delta_i\]
which contradicts the fact that $\Delta_i$ is the restricted sensitivity of ${\cal T}_i$.
\end{proof}

\subsection{Efficient Procedures for $\HH_k$ via Projection Schemes}
%Combining Projection Schemes with Restricted Sensitivity} 

%Our first result shows that given any hypothesis and any query $f$ it is possible to construct a query $f_{\cal H}$ whose global sensitivity matches the restricted sensitivity of $f$ over ${\cal H}$. Our construction of $f_{\cal H}$ is inefficient. 

Unfortunately, the construction of Theorem \ref{thm:ConstructFprime} is highly inefficient. Furthermore, this construction deals with one query at a time. We would like to a-priori have a way to efficiently devise $f_{\cal H}$ for any $f$. In this section, the way we devise $f_{\cal H}$ is by constructing a \emph{projection} -- a function $\mu:\mathcal{D}\to\mathcal{H}$ with the property that $\mu(D) = D$ for every $D\in \mathcal{H}$. Such $\mu$ allows us to canonically convert {\em any} $f$ into $f_{\cal H}$ using the na\"ive definition $f_{\cal H} = f\circ \mu$.   Below we discuss various properties of projections that allow us to derive ``good'' $f_{\cal H}$-s. Following each property, we exhibit the existence of such projections $\mu$ for the specific case of social networks and $\HH = \HH_k$, the class of graphs of degree at most $k$.
%%, projected onto the set of graphs of bounded-degree.

\begin{definition}
\label{def:Hk}
The class $\mathcal{H}_k$ is defined as the set $\{G\in\mathcal{G}:\ \forall v,~\deg(v) \leq k\}$.
\end{definition}
In many labeled graphs, it is reasonable to believe that ${\cal H}_k$ holds for $k \ll n$ because the degree distributions follow a power law. For example, the number of telephone numbers receiving $t$ calls in a day is proportional to $1/t^2$, and the number of web pages with $t$ incoming links is proportional to $1/t^2$ \cite{newman2003structure,broder2000graph,easley2010networks}. For these networks it would suffice to set $k = O\paren{\sqrt{n}}$. The number of papers that receive $t$ citations is proportional to $1/t^3$ so we could set $k = O\paren{\sqrt[3]{n}}$ \cite{easley2010networks}. While the degrees on Facebook don't seem to follow a power law, the upper bound $k=5,000$ seems reasonable \cite{ugander2011anatomy}. By contrast, Facebook had approximately $n=901,000,000$ users in June, 2012 \cite{facebookStatistics}.

\subsubsection{Smooth Projection}
\label{sub:smooth_projection}

%To obtain efficient constructions we relax the guarantee that $GS_{f_{\cal H}} = RS_f\paren{{\cal H}}$. From now on, we are content when we devise a function $f_{{\cal H}}$ s.t. (i) $GS_{f_{\cal H}} \leq c \cdot RS_f\paren{{\cal H}}$ for some small approximation factor $c$ or (ii) $S_{f,\beta}\paren{G} \leq c \cdot RS_f\paren{{\cal H}}$ for some constant $c$ whenever $G \in {\cal H}$. In the second case we also need to guarantee that the $\beta$-smooth upper bound $S_{f,\beta}$ is efficiently computable. We outline several sufficient conditions to achieve these goals.

The first property we discuss is perhaps the simplest and most coveted property such projection can have -- smoothness. Smoothness dictates that there exists a global bound on the distance between any two mappings of two neighboring databases.

\begin{definition}
\label{def:smooth_projection}
A projection $\mu:\mathcal{D}\to\mathcal{H}$ is called \emph{$c$-smooth} if for any two neighboring databases $D\sim D'$ we have that $d\big(\mu(D),\mu(D')\big)\leq c$.
\end{definition}

\begin{lemma} \label{lemma:MappingLemma1}
Let $\mu: \mathcal{D} \rightarrow {\cal H}$ be a $c$-smooth projection (i.e., for every $D\in \mathcal{H}$ we have $\mu(D) = D$). Then for every query $f$, the function $f_{\cal H} = f\circ \mu$ satisfies that $GS_{f_{\cal H}} \leq c \cdot RS_f\paren{{\cal H}} \ . $
\end{lemma}
\begin{proof}
\begin{eqnarray*}
& GS_{f_{{\cal H}}}  &= \max_{D_1\sim D_2} {\left| f_{{\cal H}}\paren{D_1} - f_{{\cal H}}\paren{D_2}\right| } \cr
&& =  \max_{D_1\sim D_2}{\left| f\paren{\mu\paren{D_1}} - f\paren{\mu\paren{D_2}}\right|}\cdot 1  \cr
&&\leq \max_{D_1\sim D_2} {\left| f\paren{\mu\paren{D_1}} - f\paren{\mu\paren{D_2}}\right|}~\frac c {d\paren{\mu\paren{D_1},\mu\paren{D_2}}}  \cr
&&\leq c \cdot \max_{D_1,D_2 \in {\cal H}} \frac{\left| f\paren{D_1} - f\paren{D_2}\right| }{d\paren{D_1,D_2}} \cr
&& = c\cdot RS_f \paren{{\cal H}} 
\end{eqnarray*}
\end{proof}

\cut{
\begin{remark}
In fact finding a suitable value of $f_{\cal H}$ is equivalent solving the following LP
\[f_{\cal H}\paren{D_{i+1}}= \min x \] such that
\[ \forall D \in {\cal H}, x \geq f\paren{D} - RS_f\paren{{\cal H}} d\paren{D,D_{i+1}} \ .\]
From an algorithmic standpoint notice that $f'$ can be computed independently for each $D_{i+1} \notin {\cal H}$. In some special cases the LP could be solved efficiently. For example if there is a separation oracle.  
\end{remark}}

As we now show, for $\HH = \HH_k$ and for distances defined via the edge-adjacency model, we can devise an efficient smooth projection. 

%The key idea behind theorem \ref{thm:ConstructFprimeEfficient} is to define an efficient mapping $\mu:\mathcal{G}\rightarrow G$ which satisfies 1) $\mu\paren{G} = G$ for all $G \in {\cal H}$ and 2) $d\paren{\mu\paren{G_1},\mu\paren{G_2}} \leq 3 \cdot d\paren{G_1,G_2}$, and set $f_{{\cal H}_k}\paren{G} = f\paren{\mu\paren{G}}$. The mapping $\mu$ works by fixing a cannonical ordering over all edges. An edge $e$ is deleted if and only if there is a vertex $v$ such that 1) $e$ is incident to $v$ and 2) $e$ is not one of the first $k$ edges incident to $v$. In the appendix we prove that this mapping satisfies conditions 1) and 2). The result follows from lemma \ref{lemma:MappingLemma1}.

\begin{claim}
\label{clm:edge-adjacency-projection}
%In the edge-adjacency model, there exists an efficient way to compute a $3$-smooth projection $\mu$ to $\HH_k$.
In the edge-adjacency model, there exists an efficiently computable $3$-smooth projection to $\HH_k$.
\end{claim}
The proof of the claim is deferred to the appendix. The high-level idea is to fix a canonical ordering over all edges and then define $\mu$ to delete an edge $e$ if and only if there is a vertex $v$ such that (1) $e$ is incident to $v$ and (2) $e$ is not one of the first $k$ edges incident to $v$.  This is then used to achieve the smoothness guarantee.  
An immediate corollary of Lemma~\ref{lemma:MappingLemma1} and Claim~\ref{clm:edge-adjacency-projection} is the following theorem.

\begin{theorem} \label{thm:ConstructFprimeEfficient}
(Privacy wrt Edge Changes) Given any query for social networks $f$, the mechanism that uses the projection $\mu$ from Claim~\ref{clm:edge-adjacency-projection}, and answers the query using $A(f,G) = f(\mu(G)) + Lap(3\cdot RS_f({\cal H}_k)/\epsilon)$ preserves $(\epsilon,0)$ privacy for any graph $G$.
\end{theorem}

Now, it is evident that this mechanism has the guarantee that for every $G\in\mathcal{H}_k$ it holds that $\Pr[ ~ |A(f,G) - f(G)| \leq O(RS_f({\cal H}_k)/\epsilon)] \geq 2/3$. Furthermore, if the querier ``lucked out'' to ask a query $f$ for which $f(G)$ and $f(\mu(G))$ are close (say, identical), then the same guarantee holds for such $G$ as well. Note however that we \emph{cannot reveal} to the querier whether $f(G)$ and $f(\mu(G))$ are indeed close, as such information might leak privacy.

\subsubsection{Projections and Smooth Distances Estimators}
\label{subsec:smooth-distance-estimators}

Unfortunately, the smooth projections do not always exist, as the following toy-example demonstrates. Fix $n$ graphs, where $d\paren{G_i,G_j} = \left|i-j\right|$ for $1 \leq i,j \leq n$, and let ${\cal H} = \{G_1, G_n\}$. Because $\mu\paren{G_1} = G_1$ and $\mu\paren{G_n}=G_n$, then there must exist some value $i$ such that $\mu\paren{G_i} \neq \mu\paren{G_{i+1}}$, thus every $\mu$ cannot be $c$-smooth for $c<n$. 

Note that smooth projections have the property that they also provide a $c$-approximation of the distance of $D$ to $\mathcal{H}$. Meaning, for every $D$ we have that $d(D,{\cal H}) \leq d(D,\mu(D)) \leq c \cdot d(D,{\cal H})$. In the vertex adjacency model, however, it is evident that we cannot have a $O(1)$-smooth projection since, as we show in the appendix, it is NP-hard to approximate $d(G, {\cal H}_k)$ (see Claim~\ref{apxclaim:mappingHardness}), but there does exists an efficient approximation scheme (see Claim~\ref{apxclaim:approxMapping}) of the distance. Yet, we show that it is possible to devise a somewhat relaxed projection s.t. the distance between a database and its mapped image is a smooth function. To that end, we relax a little the definition of projection, allowing it to map instances to some predefined $\bar{\mathcal{H}} \supset \mathcal{H}$.

\begin{definition}
\label{dfn:smooth-distance}
Fix $\bar{\mathcal{H}} \supset \mathcal{H}$. Let $\mu$ be a projection of $\mathcal{H}$, so $\mu$ is a mapping $\mu: \mathcal{D} \to \bar{\mathcal{H}}$ that maps every element of ${\cal H}$ to itself ($\forall D\in\mathcal{H}$ we have that $\mu(D) = D$). A \emph{$c$-smooth distance estimator} is a function $\hat d_\mu:\mathcal{D} \to\mathbb{R}$ that satisfies all of the following. (1) For every $D\in\mathcal{H}$ it is defined as $\hat d_\mu(D) = 0$. (2) It is lower bounded by the distance of $D$ to its projection: $\forall D\in \mathcal{D}, \ \hat d_\mu(D) \geq d(D,\mu(D))$. (3) Its value over neighboring databases changes by at most $c$: $\forall D\sim D', \ \abs{ \hat d_\mu(D) - \hat d_\mu(D')  } \leq c$.
\end{definition}
It is simple to verify that for every $D\in\mathcal{D}$ we have that $\hat d_\mu(D) \leq c \cdot d(D,\mathcal{H})$ (using induction on $d(D,\mathcal{H})$). We omit the subscript when $\mu$ is specified. 

The following lemma suggests that a smooth distance estimator allows us to devise a good smooth-upper bound on the local-sensitivity, thus allowing us to apply the smooth-sensitivity scheme of~\cite{nissim2007smooth}.

%Lemma \ref{lemma:mappingLemma2} below suggests an alternative approach. The key idea is that we can allow the local sensitivity at $D$ to grow with $d\paren{D,{\cal H}}$ (e.g., $LS_{f_{\cal H}} \paren{D} \leq \paren{ 2 \cdot d\paren{D}+1} RS_f$) without adversly affecting the smooth sensitivity of datasets $D \in {\cal H}$.

\begin{lemma}\label{lemma:mappingLemma2}
Fix $\bar{\mathcal{H}}\supset \mathcal{H}$ and let $\mu:{\cal D}\to\bar{\cal H}$ be a projection of $\mathcal{H}$. Let $\hat{d}:\mathcal{D}\rightarrow \mathbb{R}$ be an efficiently computable $c$-smooth distance estimator. Then for every query $f$, we can define the composition $f_{\cal H} = f\circ \mu$ and define the function \[S_{f_{\cal H}, \beta}\paren{D} = \max_{d \in \mathbb{Z}, d \geq \hat{d}\paren{D}} e^{\paren{-\tfrac \beta c \paren{d-\hat{d}\paren{D}} } }\paren{2 d + c+1 }  \cdot RS_{f} \paren{\bar{\cal H}} \  \]
Then $S_{f_{\cal H}, \beta}$ is an efficiently computable $\beta$-smooth upper bound on the local sensitivity of ${f_{\cal H}}$. Furthermore, define $g$ as the function $ g(x) = \begin{cases} 2 \tfrac 1 x e^{-1 + \tfrac {c+1} 2 x}, & 0\leq x\leq \tfrac 2{c+1} \cr c+1, & x> \tfrac 2{c+1}\end{cases} $. Then for every $D$ it holds that 
\[S_{f, \beta}\paren{D} \leq \exp(\tfrac \beta c \hat d(D)) \cdot g(\beta/c) RS_{f}(\bar{\cal H})\]
\end{lemma}

The proof of Lemma \ref{lemma:mappingLemma2} is deferred to the appendix. Like in the edge-adjacency model, we now exhibit a projection and a smooth distance estimator for the vertex-adjacency model.

\begin{claim}
\label{clm:vertex-adjacency-projection}
In the vertex-adjacency model, there exists a projection $\mu:\mathcal{G}\to\mathcal{H}_{2k}$ and a $4$-smooth distance estimator $\hat d$, both of which are efficiently computable.
\end{claim}

To construct $\mu$ and $\hat{d}$ we start with the linear program that determines a ``fractional distance'' from a graph to $\mathcal{H}_k$. This LP has $n + \binom n 2$ variables: $x_u$ which intuitively represents whether $x_u$ ought to be removed from the graph or not, and $w_{u,v}$ which represents whether the edge between $u$ and $v$ remains in the projected graph or not. We also use the notation $a_{u,v}$, where $a_{uv} = 1$ if the edge $\{u,v\}$ is in $G$; otherwise $a_{uv}=0$.
\begin{eqnarray*}
& \min \sum\nolimits_{v \in V} x_v & s.t. \\
& & (1)~\forall v,~ x_v\geq 0 \\
&& (2)~\forall u,v,~ w_{u,v} \geq 0 \\
& & (3)~ \forall u,v,~ a_{uv} \geq w_{uv} \geq a_{uv} - x_u - x_v \\
& & (4)~\forall u, \sum\nolimits_{v\neq u} w_{u,v} \leq k \\
\end{eqnarray*}
To convert our fractional solution $(\bar x^*,\bar w^*)$ to a graph $\mu\paren{G} \in {\cal H}_{2k}$ we define $\mu\paren{G}$ to be the graph we get by removing every edge $(u,v) \in E(G)$ whose either endpoint has weight $x^*_u > 1/4$ or $x^*_v \geq 1/4$. We define our distance estimator as
%we set $x_v^* = 1$ whenever $x_v \geq 1/4$ (otherwise $x_v^*=x_v$) and define 
$\hat{d}\paren{G} = 4 \sum_u x^*_u$. In the appendix we show that $\mu$ and $\hat{d}$ satisfy the conditions of claim \ref{clm:vertex-adjacency-projection}.

As before, combining Lemma~\ref{lemma:mappingLemma2} with Claim~\ref{clm:vertex-adjacency-projection} gives the following theorem as an immediate corollary.

\begin{theorem} \label{thm:main} 
(Privacy wrt Vertex Adjacency) Given any query for social networks $f$, the mechanism that uses the projection $\mu$ from Claim~\ref{clm:edge-adjacency-projection} and the $\beta$-smooth upper bound of Lemma~\ref{lemma:mappingLemma2}, and answers the query using $A(f,G) = f(\mu(G)) + Lap(2\cdot S_{f_{\cal H}, -\epsilon/2\ln\delta}(G)/\epsilon)$ preserves $(\epsilon,\delta)$ privacy for any graph $G$.
\end{theorem}

Again, it is evident from the definition that the algorithm has the guarantee that for every $G\in\mathcal{H}_k$ it holds that $\Pr[ ~ |A(f,G) - f(G)| \leq O(g(\tfrac \epsilon {8\ln(1/\delta)})RS_f({\cal H}_{2k})/\epsilon)] \geq 2/3$.

\section{Restricted Sensitivity and ${\cal H}_k$} \label{sec:RestrictedSensitivity}
Now that we have constructed the machinery of restricted sensitivity, we compare the restricted sensitivity over ${\cal H}_k$ with smooth sensitivity for specific types of queries, in order to demonstrate the benefits of our approach. In a nutshell, restricted sensitivity offers a significant advantage over smooth sensitivity whenever $k \ll n$. I.e., we show that there are queries $f$ s.t. for some $G \in {\cal H}_k$ it holds that $RS_f\paren{{\cal H}_k} \ll S_{f,\beta}\paren{G}$.

We now define two types of queries. First, let us introduce some notation. A profile is a function that maps a vertex $v$ in a social network $\paren{G,\ell}$ to $\left[0,1\right]$. Given a set of vertices $\{v_1, v_2, \ldots, v_t\}$, we denote by $G[v_1, v_2, \ldots, v_t]$ the social network derived by restricting $G$ and $\ell$ to these $t$ vertices. We use $G_v = G[\{v\} \cup \{ w ~\vline \paren{v,w} \in E\paren{G} \}]$ to denote the social network derived by restricting $G$ and $\ell$ to $v$ and its neighbors. A {\em local} profile satisfies the constraint $p\paren{v,\paren{G,\ell}} = p\paren{v, G_v}$.

\begin{definition} \label{dfn_query_NodesWithProfile} 
A {\em (local) profile query} \[f_p\left(G,\ell\right) = \sum\nolimits_{v \in V(G)} p\big(v,\paren{G,\ell}\big)\] sums the (local) profile  $p$ accross all  nodes.
\end{definition}
Local profile queries are a natural extension of predicates to social networks, which can be used to study many interesting properties of a social network like clustering coefficients\cite{watts1998collective,newman2003structure,bearman2004suicide}, local bridges \cite{easley2010networks, granovetter1995getting} and $2$-betweeness \cite{freeman1979centrality}. Further dissussion can be found in section \ref{sec:LocalProfileQueries} in the appendix. Claim \ref{claim:RestrictedSensitivityLocalProfileQuery} bounds the restricted sensitivity of a local profile query over ${\cal H}_k$ (e.g., in the vertex adjacency model a node $v$ can at worst affect the local profiles of itself, its $k$ old neighbors and its $k$ new neighbors). A formal proof of Claim \ref{claim:RestrictedSensitivityLocalProfileQuery} is deferred to the appendix.

\begin{claim} \label{claim:RestrictedSensitivityLocalProfileQuery}
For any local profile query $f$, we have that $RS_f\paren{{\cal H}_k} \leq 2k+1$ 
in the vertex adjacency model, and $RS_f\paren{{\cal H}_k} \leq k+1$ in the edge adjacency model. 
\end{claim}

By contrast the smooth sensitivity of a local profile query may be as large as $O(n)$ even for graphs in ${\cal H}_k$. Consider the local profile query ``how many people are friends with a spy?'' The $n-1$-star graph $G_1$ in which a spy $v$ is friends with everyone is adjacent to the empty graph $G_0 \in \HH_k$.  Therefore, any smooth upper bound $S_{f,\beta}$ must have $S_{f,\beta}(G) \geq n-1$. It is also worth observing that the assumption $G \in{\cal H}_k$ does not necessarily shrink the range of possible answers to a local profile query $f$  (e.g., there are graphs $G \in {\cal H}_k$ in which everyone is friends with a spy). 

Subgraph queries allows us to ask questions such as ``how many triplets of people are all friends when two of them are doctors and the other is a pop-singer?'' or ``how many paths of length 2 are there connecting a spy and a pop-singer over $40$?'' The average clustering coefficient of a graph can be computed from the number of triangles and $2$-stars in a graph. 

\begin{definition}\label{dfn_query_SubgraphCounting} 
A \emph{subgraph counting} query $f = \langle H, \bar p\rangle$ is given by a connected graph $H$ over $t$ vertices and $t$ predicates $p_1, p_2, \ldots, p_t$. Given a social network $(G,\ell)$, the answer to $f\paren{G,\ell}$ is the size of the set \[\big\{ v_1, v_2, \ldots, v_t : \ G[v_1,v_2, \ldots,v_t] = H \textrm{ and  } ~ \forall i, ~ \ell\paren{v_i} \in p_i \big\} \]
\end{definition}

The smooth sensitivity of a subgraph counting query may be as high as $O\paren{n^{t -1}}$ in the vertex adjacency model. Let $f = \langle H, \bar{p}\rangle$ be a subgraph counting query where $H$ is a $t$-star and each predicate $p_i$ is identically true. Let $G_1$ be a $n$-star ($f\paren{G_1} = \binom{n}{c-1}$). Then in the vertex adjacency model there is a neighboring graph $G_2$ with no edges ($f\paren{G_2} = 0$). We have that $LS_{f}\paren{G_2} \geq  \binom{n}{t-1}$. Observe that $G_2 \in {\cal H}_k$. In the appendix we show that the smooth sensitivity  of $f = \langle {K_3}, \bar{p} \rangle$ is {\em always} greater than $n$ when each predicate $p_i$ is identically true (see claim \ref{claim:SmoothSensitivitySubgraphCounting}).  By contrast Claim \ref{claim:RestrictedSensitivitySubgraphCounting} bounds the restricted sensitivity of subgraph counting queries. The proof is deferred to the appendix.

\begin{claim} \label{claim:RestrictedSensitivitySubgraphCounting}
Let $f = \langle H, \bar p\rangle$ be subgraph counting query and let $t = \left| H \right|$ then $ RS_{f}\paren{{\cal H}_k} \leq t k^{t-1} $
in the edge adjacency model and in the vertex adjacency model. 
\end{claim}

While the assumption $G \in {\cal H}_k$ may shrink the range of a subgraph counting query $f$, the restricted sensitivity of $f$ will typically be much smaller than this reduced range. For example, if $f(G)$ counts the number of triangles in $G$ then $f(G) \leq nk^2$ for any $G \in {\cal H}_k$, while $RS_{f}\paren{{\cal H}_k} \leq 3 k^2 \ll nk^2$.

\section{Future Questions/Directions} \label{sec:futureQuestions}
{\bf Efficient Mappings:} While we can show that there doesn't exist an efficiently computable $O(1)$-smooth projection $\mu:\mathcal{G} \rightarrow \HH_k$, we don't know whether the construction of Claim~\ref{clm:vertex-adjacency-projection} can be improved. Meaning, there could be a mapping $\mu:\mathcal{G}\to\bar{\HH}$ for some $\bar{\HH}\supset \HH_k$, whether the solution itself, the set of vertices that dominate the removed edges, is smooth. In other words, 
Is there an efficiently computable mapping $\mu:\mathcal{G} \rightarrow \bar{\HH} \subset \HH_k$ which satisfies $\left| d\paren{\mu\paren{G_1},G_1}-d\paren{\mu\paren{G_2},G_2} \right| \leq c$ for some constant $c$? {\bf Multiple Queries: } We primarily focus on improving the accuracy of a single query $f$. Could the notion of restricted sensitivity be used in conjunction with other mechanisms (e.g., BLR \cite{Blum:2008:LTA:1374376.1374464}, Private Multiplicative Weights mechanism \cite{hardt2010multiplicative}, etc.) to accurately answer an entire class of queries? {\bf Alternate Hypotheses:} We focused on the specific hypothesis $\HH_k$. What other natural hypthothesis could be used to restrict sensitivity in private data analysis? Given such a hypothesis $\HH$ can we efficiently construct a query $f_H$ with low global sensitivity or with low smooth sensitivity over datasets $D \in \HH$?

\bibliographystyle{plain}
\bibliography{socialNetworks}

\appendix
\section{Missing Proofs}
\label{apx_sex:proofs}

\begin{reminderclaim}{\ref{clm:edge-adjacency-projection}}
In the edge-adjacency model, there exists an efficient way to compute a $3$-smooth projection $\mu$ to $\HH_k$.
\end{reminderclaim}

\begin{proofof}{Claim~\ref{clm:edge-adjacency-projection}}
We construct our smooth-projection $\mu$ by first fixing a canonical ordering over all possible edges. Let $e^v_1,...,e^v_t$ denote the edges incident to $v$ in canonical order. For each edge $e = \{u,v\}$ we delete $e$ if and only if (i)  $e = e^v_j$ for $j > k$ or (ii)  $e = e^u_j$ for $j > k$ (Intuitively for each $v$ with $\deg\paren{v} \geq k$ we keep this first $k$ edges incident to $v$ and flag the other edges for deletion). If $G \in {\cal H}_k$ then no edges are deleted, so $\mu(G) = G$. Suppose that $G_1, G_2$ are neighbors differing on one edge $e = \{x,y\}$ (wlog, say that $e$ is in $G_1$). Observe that for every $v \neq x,y$, the same set of edges incident to $v$ will be deleted from both $G_1$ and $G_2$. In fact, if $\mu(G_1)$ does not contain $e$ then $\mu(G_1) = \mu(G_2)$. Otherwise, if $e$ is not deleted we may assume then there may be at most one edge $e_x$ (incident to $x$) and at most one edge $e_y$ (incident to $y$) that were deleted from $\mu\paren{G_1}$ but not from $\mu{\paren{G_2}}$. Hence, $d\paren{\mu\paren{G_1},\mu\paren{G_2}} \leq 3$. 
\end{proofof}

\begin{reminderlemma}{\ref{lemma:mappingLemma2}}
Fix $\bar{\mathcal{H}}\supset \mathcal{H}$ and let $\mu:{\cal D}\to\bar{\cal H}$ be a projection of $\mathcal{H}$. Let $\hat{d}:\mathcal{D}\rightarrow \mathbb{R}$ be an efficiently computable $c$-smooth distance estimator. Then for every query $f$, we can define the composition $f_{\cal H} = f\circ \mu$ and define the function \[S_{f_{\cal H}, \beta}\paren{D} = \max_{d \in \mathbb{Z}, d \geq \hat{d}\paren{D}} e^{\paren{-\tfrac \beta c \paren{d-\hat{d}\paren{D}} }} \paren{2 d + c+1 }  \cdot RS_{f} \paren{\bar{\cal H}} \  \]
Then $S_{f_{\cal H}, \beta}$ is an efficiently computable $\beta$-smooth upper bound on the local sensitivity of ${f_{\cal H}}$. Furthermore, define $g$ as the function $ g(x) = \begin{cases} 2 \tfrac 1 x e^{-1 + \tfrac {c+1} 2 x}, & 0\leq x\leq \tfrac 2{c+1} \cr c+1, & x> \tfrac 2{c+1}\end{cases} $. Then for every $D$ it holds that 
\[S_{f, \beta}\paren{D} \leq \exp(\tfrac \beta c \hat d(D)) \cdot g(\beta/c) RS_{f}(\bar{\cal H})\]
\end{reminderlemma}
\begin{proofof}{Lemma \ref{lemma:mappingLemma2}}
First, we show that indeed $S_{f_{\cal H}, \beta}$ is an upper bound on the local sensitivity of $f_{\cal H}$. Fix any $D\in \mathcal{D}$ and indeed
\begin{eqnarray*}
& LS_{f_{\cal H}}\paren{D} &=  \max_{D' \sim D} \left| f_{\cal H}\paren{D} - f_{\cal H}\paren{D'} \right| \\
&& =  \max_{D' \sim D} \left| f(\mu(D))-f(\mu(D')) \right| \\
& & \leq \max_{D'\sim D} RS_f(\bar{\mathcal{H}}) \cdot d(\mu(D),\mu(D')) \\ 
& & \leq \max_{D'\sim D} RS_f(\bar{\mathcal{H}}) \cdot \\ 
&& \qquad  \big( d(D,\mu(D)) + d(D,D') + d(D',\mu(D')) \big)\\ 
& & \leq RS_f(\bar{\mathcal{H}}) \cdot \big( \hat d(D)+1+ \max_{D'\sim D} \hat d(D') \big)\\
&&  \leq RS_f(\bar{\mathcal{H}}) \cdot \big( 2\hat d(D)+c+1\big)\\
& & \leq \max_{d \geq \hat{d}\paren{D} } e^{-\beta c \paren{d - \hat{d}\paren{D} }}  \paren{2d+c+1} RS_f\paren{\bar{\cal H}} \\
& & = S_{f_{\cal H}, \beta}\paren{D} \ . 
\end{eqnarray*}  

Next we prove that $S_{f_{\cal H}, \beta}$ is $\beta$-smooth. Let $D_1$ and $D_2$ be two neighboring databases, and wlog assume $\hat d(D_2) \leq \hat d(D_1)$. Then
\begin{eqnarray*}
&& \frac {\displaystyle S_{f_{\cal H}, \beta}(D_1)}{\displaystyle S_{f_{\cal H}, \beta}(D_2)} \\
&& = \frac{\max_{d \geq \hat{d}\paren{D_1}} e^{\paren{-\tfrac \beta c \paren{d-\hat{d}\paren{D_1}} }} \paren{2 d+ c + 1 } RS_f\paren{\bar{\cal H}}}{\max_{d \geq \hat{d}\paren{D_2}} e^{\paren{-\tfrac \beta c \paren{d-\hat{d}\paren{D_2}} }} \paren{2 d+ c+1 } RS_f\paren{\bar{\cal H}}}
\end{eqnarray*}
Let $d_0$ be the value of $d$ on which the maximum of numerator is obtained. Then
\begin{eqnarray*}
&& \frac {\displaystyle S_{f_{\cal H}, \beta}(D_1)}{\displaystyle S_{f_{\cal H}, \beta}(D_2)} \\ 
&& = \frac{\exp\paren{-\tfrac \beta c \paren{d_0-\hat{d}\paren{D_1}} } \paren{2 d_0+ c + 1 } RS_f\paren{\bar{\cal H}}}{\max_{d \geq \hat{d}\paren{D_2}} \exp\paren{-\tfrac \beta c \paren{d-\hat{d}\paren{D_2}} } \paren{2 d+ c+1 } RS_f\paren{\bar{\cal H}}} \\
&& \leq \frac{ \exp\paren{-\tfrac \beta c \paren{d_0-\hat{d}\paren{D_1}} } \paren{2 d_0+ c + 1 } RS_f\paren{\bar{\cal H}}}{\exp\paren{-\tfrac \beta c \paren{d_0-\hat{d}\paren{D_2}} } \paren{2 d_0+ c+1 } RS_f\paren{\bar{\cal H}}}\\
&& = \exp\left( -\tfrac \beta c (\hat d(D_2)-\hat d(D_1)) \right)  \leq \exp(\beta)
\end{eqnarray*}
where the last inequality uses the smoothness property, i.e. that $\hat d(D_2) - \hat d(D_1) \geq -c$.

Finally, we wish to prove the global upper bound on $S_{f_{\cal H}, \beta}$, i.e., that for every $D \in {\cal D}$
\[S_{f_{\cal H},\beta}\paren{D} \leq \exp(\tfrac \beta c\hat d(D)) \cdot  g\paren{c/\beta}  RS_f\paren{\bar{\cal H}} \ . \]
Fix $D$ and define $h(x) = \exp\big(-\tfrac \beta c x\big)(2x+c+1)$, so that $S_{f_{\cal H},\beta} = \exp\big(\tfrac \beta c \hat d(D)\big)RS_f(\bar{\cal H}) \cdot \max\limits_{d\geq d_0} h(d)$. Taking the derivative of $h$ we have 
\[ h'(x) = e^{-\tfrac \beta c x} \left(- 2x\tfrac \beta c - \beta -\tfrac \beta c+2 \right) \]
which means that $h(x)$ is maximized at $x_0 = \tfrac c \beta - \tfrac {c+1} 2$. In the case that $x_0 < 0$ (i.e. for $\beta / c > \tfrac 2 {c+1}$) we can upper bound the function $h(x)$ with $h(0)=c+1$ for every $x\geq 0$. Otherwise, we have that $h(x) \leq h(x_0)$ for every $x \geq 0$, and indeed $h(x_0) = 2\tfrac  c \beta e^{-1 + \tfrac \beta c \cdot \tfrac {c+1} 2} = g(\beta/c)$.

To conclude the proof, observe that computing $S_{f_{\cal H}, \beta}\paren{D}$ is just a simple optimization once $\hat{d}\paren{D}$ is known, much like the derivation done above. So since $\hat{d}$ is efficiently computable, we have that $S_{f_{\cal H}, \beta}$ is efficiently computable. \end{proofof}

\begin{reminderclaim}
{\ref{clm:vertex-adjacency-projection}}
In the vertex-adjacency model, there exists a projection $\mu:\mathcal{G}\to\mathcal{H}_{2k}$ and a $4$-smooth distance estimator $\hat d$, both of which are efficiently computable.
\end{reminderclaim}

\begin{proofof}{{Claim~\ref{clm:vertex-adjacency-projection}}}
We first prove that $\mu$ is a projection mapping every graph to a graph in $\mathcal{H}_{2k}$. Suppose that some $v \in G$ has degree $\geq 2k$, then clearly $x^*_v \leq 1/4$, for otherwise we would have removed all of the edges touching $v$. Observe that every edge we keep has $w^*_{u,v} \geq 1 - 1/4 - 1/4 =  1/2$. Consequently, we can have at most $2k$ edges with $w_{u,v} \geq \frac{1}{2}$ because of the constraint $\sum_{u} w_{u,v} \leq k$. So there are at most $2k$ edges incident to $v$ in $\mu\paren{G}$. 

Now, let us prove that $\hat d$ satisfies all of the requirements of a $4$-smooth distance estimator. First, if $G\in \mathcal{H}_k$ then the optimal solution of the LP is the all zero vector, so $\hat d(G) = 0$ for all graphs of max-degree $\leq k$. Secondly, observe that in the process of computing $\mu(G)$, every edge that is removed from $G$ can be ``charged'' to a vertex $v$ with $x_v^*\geq 1/4$. If follows that 
\[d(G,\mu(G)) \leq \sum\limits_{v: x_v^* \geq 1/4} 1 \ \ \leq \sum\limits_{v: x_v^* \geq 1/4} 4x_v^* \ \ \leq 4\sum_v x_v^* \ \ = \hat d(G)\ .\]
Lastly, fix any neighboring $G_1, G_2\in\mathcal{G}$, and let $v$ be the vertex whose edges differ in $G_1$ and $G_2$. Clearly, if $\bar x^*$ is a solution for $LP(G_1)$, then we set $y_{v} = 1$ for $i=1...d$ and $y_{v} = x^*_v$ otherwise. Now $\vec{y}$ is a feasible (not necessarily optimal) solution to $LP\paren{G_2}$. It is simple to infer that
\begin{eqnarray*}
& \hat{d}\paren{G_2}-\hat{d}\paren{G_1} &=  \hat{d}\paren{G_2}- 4 \sum_u x^*_u  \\
&& \leq  4 \sum_u y_u - 4 \sum_u x^*_u \leq 4 \sum_{u} \left| \paren{ y_u - x^*_u} \right| \\
&&= 4 \left| y_{v} - x^*_{v}\right| \leq 4 
\end{eqnarray*}
\end{proofof}

\begin{reminderclaim}{ \ref{claim:RestrictedSensitivityLocalProfileQuery}}
For any local profile query $f$, we have that $RS_f\paren{{\cal H}_k} \leq 2k+1$ 
in the vertex adjacency model and $RS_f\paren{{\cal H}_k} \leq k+1$ in the edge adjacency model. 
\end{reminderclaim}

\begin{proofof}{ Claim \ref{claim:RestrictedSensitivityLocalProfileQuery}}
Consider a local profile query $f_p$.\\ 
(Label change) Let $G_1, G_2 \in {\cal H}$ be two graphs with the same exact edge set, but with labeling functions $\ell_1, \ell_2$  that are different on a single vertex. Let $v$ be the vertex whose label differs on $G_1$ and $G_1$, and let $N_v$ denote the set of its (at most $k$) neighbors. Then for every $u \notin \{v\} \cup N_v$ we have that $p\paren{u,\paren{G_1,\ell_1}} = p\paren{u,\paren{G_2,\ell_2}}$. Hence, $\left|f_p\paren{G_1}-f_p\paren{G_2} \right| \leq | \{v\} \cup N_v | \leq k+1$.\\
(Vertex Adjacency) Let $G_1, G_2 \in {\cal H}$ be any two neighboring labeled graphs such that $G_1-v = G_2-v$. Let $N^1_v$ (resp. $N^2_v$) denote the neighborhood of $v$ then for any $y \notin N^1_v \cup N^2_v$ we have that  $p\paren{y,\paren{G_1,\ell_1}} = p\paren{y,\paren{G_2,\ell_2}}$. Hence, $\left|f_p\paren{G_1}-f_p\paren{G_2} \right| \leq \left| N^1_v \cup N^2_v \cup\{v\}\right| \leq 2k+1$.\\
(Edge Adjacency)  Let $G_1, G_2 \in {\cal H}$ be any two neighboring labeled graphs. Wlog, there is an edge $e = \{u,v\}$ such that $E(G_1) = E(G_2)\cup \{e\}$. In order to have a vertex $y$ s.t. $p\paren{y,\paren{G_1,\ell_1}} \neq p\paren{y,\paren{G_2,\ell_2}}$ we need that the edge $e$ appears in graph we get by restricting the social network to set of $y$ and its neighbors. It follows that the only vertices whose local profile can change are in the union $\{u,v\} \cup \big(N_u \cap N_v \big)$. Hence, $\left|f\paren{G_1}-f\paren{G_2} \right| \leq \left| \{u\}\cup \{v\} \right| + \left| N_u \setminus \{v\}\right| \leq 2 + k - 1 = k+1$.
\end{proofof}

\begin{reminderclaim} {\ref{claim:RestrictedSensitivitySubgraphCounting}}
Let $f = \langle H, \bar p\rangle$ be subgraph counting query and let $t = \left| H \right|$ then 
\[ RS_{f}\paren{{\cal H}_k} \leq t k^{t-1} \]
in the edge adjacency model and in the vertex adjacency model. 
\end{reminderclaim}

\begin{proofof} {Claim~\ref{claim:RestrictedSensitivitySubgraphCounting}}   (Sketch) Let $G_1, G_2 \in {\cal H}_k$ be neighbors and let $v$ be a vertex such that $G_1-v = G_2-v$, and let $N_i$ denote the neighbors of $v$ in $G_i$. Any copy of $H$ which occurs in $G_1$ but not in $G_2$ must contain $v$. Because $H$ is connected we can bound the number of $G_1$ copies of $H$. We can start with $v$, and we pick one of the $t$ vertices of $H$ to be mapped to $v$. Denote this vertex as $v_0$. Now, we proceed inductively. We pick a vertex $v \in H$, connected to the set $\{v_0, v_1, \ldots, v_{i-1}\}$. The vertex $v_i$ must be assigned to a vertex in $G$ which is incident to some specific vertex of the $i$ vertices that we already mapped. Because we have bounded degree, then there are at most $k$ options from which to choose $v_i$. We obtain the bound: $f\paren{G_1} - f\paren{G_2} \leq  t \prod_{i=1}^{t-1} {k}  =  t k^{t-1} $.
\end{proofof}

\section{Additional Claims}
\label{apx_sec:additional_claims}

\begin{claim} \label{apxclaim:mappingHardness} (Privacy wrt Vertex Adjacency)
Unless $P = NP$ there is no efficiently computable mapping $\mu:\mathcal{G} \rightarrow {\cal H}_k$ such that 
\begin{enumerate}
\item $\forall G \in {\cal H}_k, \mu\paren{G} = G$.
\item $\forall G \in \mathcal{G}, d\paren{G,\mu\paren{G}} \leq O\paren{\ln \paren{k} d\paren{G,{\cal H}_k}} $.
\end{enumerate}
\end{claim}
\begin{proof} (Sketch)
Our reduction is from the minimum set cover problem. It is NP-hard to approximate the minimum set cover problem to a factor better than $O(\log n)$ \cite{raz1997sub,alon2006algorithmic}. Given a set cover instance with sets $S_1,...,S_m$ and universe $U = \{x_1,...,x_n\}$ we set $m_i = \left|\{j :\  x_i \in S_j \right|$ and $k = n+1$. We construct our labeled graph $G$ as follows:
\begin{enumerate}
\item Add a node for each $S_i$.
\item Add a node for each $x_j$. 
\item Add the edge $\{x_j,S_i\}$ if and only if $x_j \in S_i$.
\item For each $x_i$, create $k+1-m_i$ fresh nodes $y_1,...,y_{k-m_i}$ and add each edge $\{y_j,x_i\}$. 
\end{enumerate}

Intuitively each node $x_j$ has $k+1$ incident edges. By deleting all of the edges incident to the node $S_i$ we can fix all of the nodes $x \in S_i$. Hence, $d\paren{G,\paren{{\cal H}_k}}$ corresponds exactly to the size of the minimum set cover. 
\end{proof}

\begin{claim} \label{apxclaim:approxMapping} (Privacy wrt Vertex Adjacency)
There is an efficiently computable projection $\mu:\mathcal{G} \rightarrow {\cal H}_k$ such that 
for every $G \in \mathcal{G}$ it holds that $d\paren{G,\mu\paren{G}} \leq \paren{\ln \paren{2d^2+kd}} d\paren{G,{\cal H}_k} $,  
\end{claim}
\begin{proof} (Sketch) We use a greedy algorithm to create $\mu$. Define the potential of a graph $G$ as follows
\[ \phi\paren{G} = \sum_{v\in G:\deg\paren{v} \geq k} \paren{\deg\paren{v} - k}  \ .\]
Our algorithm $\mu$ starts by guessing a value $d$ for $d\paren{G,{\cal H}_k}$ and deleting any vertex with degree $\geq k+d+1$ (these vertices {\em must} be deleted because the degree will be at least $k+1$ after deleting $d$ other vertices). Then $\mu$ repeatedly picks the vertex $v$ with the highest potential and eliminates all incident edges, where the potential of a vertex $v$ is $\phi\paren{G}-\phi\paren{G-v}$. Let $\phi_i$ denote the potential after round $i$ ( $\phi_0$ is the potential after deleting vertices with degree $\geq k+d+1$). Observe that $\phi_0 \leq 2d^2+kd$ if $d$ is correct because (i) there are $d$ vertices we can delete to drop the potential to $0$ and (2) deleting a single vertex $v$ decreases the potential by at most $\deg(v) + \deg(v)-k \leq 2d+k$. Also observe that in any round there always exists some vertex whose removal decreases the potential by at least $\paren{1-1/d} \phi_i$ so we have
$\phi_i \leq \phi_{i-1}\paren{1-\frac{1}{d}}$.
Once $i \geq d \ln \paren{2d^2 + dk+d}$ we have $\phi_i \leq 1$. 
\end{proof}

The reduction in Claim~\ref{apxclaim:approxMapping} might be used to produce a function $f_{\cal H}\paren{G} = \mu \paren{f\paren{G}}$ with low smooth-sensitivity over the nice graphs ${\cal H}_k$. Unfortunately, we don't know of any efficient algorithm to compute the smooth upper bound for such $f_{\cal H}$.

\begin{claim} \label{claim:SmoothSensitivitySubgraphCounting}
Let $f = \langle K_3, \bar{p} \rangle$ be a subgraph counting query with predicates $p_i$ that are identically true. In the vertex adjacency model for any $\beta$ smooth upper bound on the local sensitivity of $f$ and {\em any} graph $G$ we have \[S^*_{f^P,\beta}\left(G\right) \geq \exp\left(-2\beta \right)\left(n-2\right) \ . \]
\end{claim}
\begin{proof}
Let $G$ be given. Pick $v_1,v_2 \in V(G)$ and let $G_1$ be obtained from $G$ by adding all possible edges incident to $v_1$ and let $G_2$ be obtained from $G_1$ by  deleting all edges incident to $v_2$. Finally, let $G_3$ be obtained from $G_2$ by adding all possible edges incident to $v_2$.  Now the local sensitivity of $f$ at $G_2$ is at least $n-2$,
\begin{eqnarray*}
&LS_{f}\left(G_2\right) & = \max_{G':d\paren{G_2,G'}=1} \left| f\left(G_2\right)-f\left(G'\right)\right| \\
&& \geq f\left(G_3\right)-f\left(G_2\right) \geq n-2  \end{eqnarray*}
Plugging this lower bound into the definition of $\beta$ smooth sensitivity we obtain the required result:
\[S^*_{f^P,\beta}\left( G \right) \geq e^{-\beta d\left(G,G_2 \right) } LS_{f^P}\left(G_2\right) \geq e^{-2\beta} \left(n-2\right) \]
\end{proof}

\section{Local Profile Queries} \label{sec:LocalProfileQueries}
Local profile queries are a natural extension of predicates to social networks, which can be used to study many interesting properties of a social network like clustering coefficients, local bridges and $2$-betweeness).
 The clustering coefficient $c(v)$ \cite{watts1998collective,newman2003structure} of a node $v$ (e.g., the probability that two randomly selected friends of $v$ are friends with each other) \vfill\eject  has been used to identify teenage girls who are more likely to consider suicide\cite{bearman2004suicide}. One explanation, is that it becomes an inherent source of stress if a person has many friends who are not friends with each other \cite{easley2010networks}. Observe that $c(v)$ {\em is} a local profile query.  An edge $\{v,w\}$ is a local bridge if its endpoints have no friends in common. A local profile could score a vertex $v$ based on the number local bridges incident to $v$. A marketing agency may be interested in identifying nodes that are incident to many local bridges because local bridges ``provide their endpoints with access to parts of the network - and hence sources of information - that they would otherwise be far away from \cite{easley2010networks}." For example, a 1995 study showed that the best job leads often come from aquaintances rather than close friends \cite{granovetter1995getting}. $2$-betweeness (a variant of betweeness \cite{freeman1979centrality}) measures the centrality of a node. We say that the $2$-betweeness of a vertex $v$ is the probability that the a randomly chosen shortest path between two randomly chosen neighbors of $v$ $x,y \in G_v$ goes through $v$.

\end{document}